\documentclass[twocolumn,showpacs,preprintnumbers,amsmath,amssymb]{revtex4}

\usepackage{graphicx}
\usepackage{dcolumn}
\usepackage{bm}
\usepackage{epstopdf}

\begin{document}

\preprint{APS/123-QED}

\title{Possibility for $J/\Psi$ suppression in high multiplicity proton-proton collisions at $\sqrt{s_{NN}}=7\,$TeV}

\author{Thomas Lang}
\author{Marcus Bleicher}%
\affiliation{%
Frankfurt Institute for Advanced Studies (FIAS) and\\
Institut f\"ur Theoretische Physik, Johann Wolfgang Goethe-Universit\"at\\
Max-von-Laue-Str. 1, 60438 Frankfurt am Main, Germany
}%

\date{\today}

\begin{abstract}
We study $J/\Psi$ absorption in high multiplicity proton-proton (pp) collisions at $\sqrt
{s_{NN}}=7\,$TeV. We predict a modification of the $J/\Psi$-yield 
within the UrQMD transport approach, where explicit
interactions of the $J/\Psi$ with the surrounding comovers and a prehadronic phase 
with adjusted cross sections and $J/\Psi$ melting is included. 
We present the analog of the nuclear modification factor in pp collisions 
at various charged particle multiplicities. It turns out that $J/\Psi$-Mesons may be 
suppressed towards higher particle multiplicities in pp collisions at LHC energies. 
\end{abstract}

\pacs{Valid PACS appear here}
\maketitle

\section{Introduction}
\markright{Thomas Lang and Marcus Bleicher. Possible $J/\Psi$ suppression in high multiplicity proton-proton collisions at LHC energies}

One major goal of ultra-high energy heavy ion physics is to recreate the phase of quarks and gluons
(the Quark Gluon Plasma, QGP) as it might have existed micro-seconds after the Big Bang.
Various experimental facilities have been built to explore the properties of the QGP experimentally, 
while on the theory side a multitude of (potential) signatures and properties of the
QGP have been predicted. Recently, the field of heavy ion physics has entered a new era with
the start of the Large Hadron Collider (LHC) at CERN. For the first time this offers the
opportunity to investigate collective effects already in proton-proton (pp) collisions, especially
when high multiplicity interactions are selected  \cite{Vogel:2010et,Werner:2010ny,Werner:2011fd,Liu:2011dk,Werner:2010ss,Vogel:2011zz}.

In this letter, we explore the absorption of $J/\Psi$-Mesons at high 
charged particle multiplicities in pp collisions at $\sqrt {s_{NN}}=7\,$TeV. 
Similar studies have recently been performed by \cite{Vogel:2010et,Werner:2010ny,Werner:2011fd,Liu:2011dk,Werner:2010ss,Vogel:2011zz} 
and found a substantial influence of the medium at high multiplicities on observables.
This study is especially interesting, because it i) allows to explore if a collectively expanding system
has been created in pp, ii) can provide insights into the pp baseline necessary for the
interpretation of heavy ion collisions and iii) provides complementary information to the studies of AA collisions.
Up to now, studies of $J/\Psi$s in a dense and hot medium were restricted to (massive)
nucleus-nucleus (AA) reactions. Here it was found that the initially produced charmonia are
suppressed in heavy ion collisions compared to pp collisions at the same energy. Three
effects to explain this suppression have been put forward:
\begin{itemize}
\item Nuclear absorption or baryonic suppression \cite{Gerschel:1998zi}. These effects depend on the
thickness function of the two nuclei and therefore change with centrality. This 'normal'
suppression with centrality is not able to describe the measured $J/\Psi$-yield in very
central AA collisions at SPS and RHIC energies \cite{Alessandro:2004ap,Arnaldi:2006it,Alessandro:2006ju,Adare:2006ns}. 
Two possible interpretations of this 'additional' suppression have been suggested and will be discussed
next.
\item Comover absorption \cite{Gavin:1988hs,Vogt:1997cs,Armesto:1997sa,Spieles:1999kp,Spieles:1998pa,Capella:2007jv}. Here the charmonia are additionally absorbed by inelastic scatterings
with comoving mesons. It is assumed that the corresponding $J/\Psi$ hadron cross
sections are on the order of a several $\textrm{mb}$ \cite{Zhang:2000nc}. Nevertheless theoretical
approximations for the absorption cross section differ by more than an order of magnitude \cite{Muller:1999ys}. In pp collisions
at low energies (i.e. low particle densities) comover absorption is irrelevant.
\item Debye screening \cite{Matsui:1986dk}. An alternative explanation of the additional suppression is explained by the formation
of a QGP. In a QGP the formation of $J/\Psi$-Mesons is suppressed due to Debye screening in
the matter. The charm quarks will leave the reaction zone as hadrons with open charm.
Therefore $J/\Psi$ suppression in heavy ion collisions has been proposed as a possible
signature for QGP formation \cite{Matsui:1986dk}. However, see also \cite{Mocsy:2008zz} for a recent discussion.
\end{itemize}
For massive nuclear collisions at higher energies, it was also speculated that $J/\Psi$-Mesons might be reformed via $J/\Psi
\leftrightarrow c\overline c$ or $J/\Psi \leftrightarrow D\overline D$ reactions, if the charm quark
(or D-Meson) densities become substantial \cite{Thews:2000rj,Andronic:2007bi,Linnyk:2007sc}. Further information in this respect might also be gained from correlation studies as suggested in \cite{Zhu:2006er,vanHees:2005wb}.

High multiplicity Proton-Proton interactions at LHC offer a completely new set of questions and possibilities 
as compared to nucleus-nucleus reactions at similar multiplicities.
At these energies the particle multiplicities at central rapidities rise to the same order of
magnitude as in heavy ion collisions at lower energies, and the energy densities may even exceed the
values of central Au-Au/Pb-Pb reactions at SPS and RHIC. This allows to investigate charmonium
suppression at high energy densities and temperatures in a rather clean environment excluding baryonic effects. 
It does also not suffer from the potential charm recombination due to the still low charm yields in pp, 
thus enabling one to explore the absorption mechanisms with higher precision.

\section{The model and results}

For the present study of $J/\Psi$-interactions with the medium, we modified the UrQMD transport approach to include 
known effects not present in the standard simulation. Here we assume a high temperature regime with melting of the 
$J/\Psi$ ($T>325\,\textrm{MeV}$), 
a prehadronic stage ($155\,\textrm{MeV} < T < 325\,\textrm{MeV}$) and finally 
a hadronic stage ($T <155\,\textrm{MeV}$). 
As UrQMD can not directly provide the temperature of the medium we map the transition 
temperatures to scalar quark densities using the chiral equation of state 
implemented in the UrQMD hydro model \cite{Petersen:2008dd}. We obtain for $T=325\,\textrm{MeV}$ 
a quark density of $n_q=12\,\textrm{fm}^{-3}$ and for $T=155\,\textrm{MeV}$ 
a quark density of $n_q=0.6\,\textrm{fm}^{-3}$. 
The local scalar quark densities at the position of the $J/\Psi$ are then used to obtain the 
corresponding $J/\Psi$ cross sections in each UrQMD time step. 
In the following the assumptions for the hadronic, prehadronic and melting phase are explained:

Melting regime ($T>325\,\textrm{MeV}$, $n_q>12/\textrm{fm}^3$):\\
The onset of charmonium melting for $J/\Psi$s is, on the basis of \cite{Petreczky:2010tk}, at $325\,\textrm{MeV}$. 
One should keep in mind that charmonium states are supposed to still persist in the QGP, their wave function should broaden nevertheless with increasing energy density \cite{Petreczky:2010tk}. 
Therefore in the present study $J/\Psi$s are assumed to melt if the energy density is high enough and the charmonia stay in this hot medium for some time. 
Here we choose, on the basis of \cite{Petreczky:2010tk}, a proper time of $1\,\textrm{fm/c}$ for the charmonium decay in the hot medium. 

\begin{figure}
\includegraphics[width=0.5\textwidth]{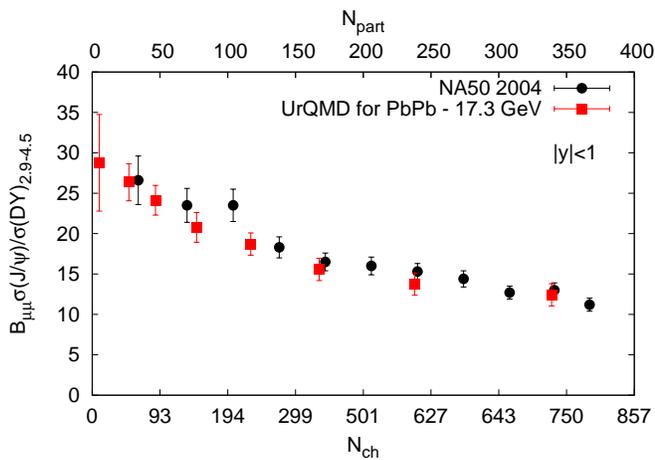}
\caption{(Color online) $J/\Psi$ decays to muons scaled by the number of Drell-Yan pairs dependent on the number of participants $N_{part}$ and the corresponding number of charged particles $N_{ch}$. 
UrQMD data are compared to data of the NA50 experiment \cite{Alessandro:2004ap}. 
The normalization factor $B_{\mu\mu}\sigma (J/\Psi)/\sigma (DY)_{2.9-4.5}\,$=34.9 in pp interactions at $158\,$GeV is taken from \cite{Alessandro:2004ap}.}
\label{SPS}
\end{figure}

Prehadronic regime
($T>155\,\textrm{MeV}$, $n_q>0.6/\textrm{fm}^3$):\\
The temperature of the phase transition is, in-line with the lattice QCD estimates \cite{Borsanyi:2010bp,Soldner:2010xk}, set to $155\,\textrm{MeV}$. 
In this prehadronic phase we mimic the QGP effects in UrQMD by pre-formed hadrons. 
To calculate charmonium dissociation we use fixed cross sections in the hot medium. The inelastic cross sections to mimic the effective dissociation of the QGP on $J/\Psi$ in the prehadronic phase 
are obtained by adjusting the hadronic cross sections to data at SPS energies, see Fig.\ref{SPS}.
The reason we use SPS data to fix the cross sections is the negligible amount of recombination of D-Mesons to charmonium states due to the low D-Meson density at SPS energies.
For this exploratory study we take constant cross sections in the prehadronic phase with a meson-$J/\Psi$ elastic and inelastic cross section of $0.78\,\textrm{mb}$, while the baryon-$J/\Psi$ cross sections are obtained by quark number scaling.

Hadronic regime ($T<155\,\textrm{MeV}$, $n_q<0.6/\textrm{fm}^3$):\\
The hadronic phase includes elastic scattering of $J/\Psi$-particles and its dissociation by baryons and mesons as well. 
Furthermore recombination of D-Mesons to charmonium states is implemented using detailed balance, but found to be negligible in pp collisions due to the low D-Meson abundance. For the inelastic charmonium-meson cross sections in the hadronic phase we use a 2-body transition model \cite{Bratkovskaya:2003ux,Bratkovskaya:2004cq}. 
\begin{equation}
\sigma_{1+2\rightarrow 3+4}(s)=2^4\frac{E_1E_2E_3E_4}{s}|M_i|^2\left(\frac{m_3+m_4}{\sqrt{s}}\right)^6\frac{p_f}{p_i}\quad.
\end{equation}
Here $E_i$ denotes the energies of the ingoing and outgoing particles, $M_i$ the masses and $p_i$ and $p_f$ the initial and final momenta in the 2-particle rest frame. 
The effective matrix element $|M_i|^2=0.65$, is fixed to Pb-Pb collisions at SPS energies. 
The matrix element is increased by a factor of three for an excited D-Meson in the outgoing channel and decreased by a factor of three for a strange D-Meson in the outgoing channel \cite{Bratkovskaya:2003ux,Bratkovskaya:2004cq}. 
The corresponding back-reaction for D-Meson recombination is given by detailed balance, 
\begin{equation}
\sigma_{3+4\rightarrow 1+2}(s)=\sigma_{1+2\rightarrow 3+4}(s)\frac{(2S_1+1)(2S_2+1)}{(2S_3+1)(2S_4+1)}\frac{p_f^2}{p_i^2}
\end{equation}
where $S_i$ denotes the spin of the particles. For the inelastic baryon-$J/\Psi$ interactions we use a constant dissociation cross section of $4.18\,\textrm{mb}$ obtained from \cite{Alessandro:2006jt,Arnaldi:2006it}.
For all elastic cross sections of charmonia in the hadronic phase we use a constant cross section of $5\,\textrm{mb}$. 
In this hadronic phase the usual UrQMD formation times are considered for the produced hadrons. 
The formation time of the $J/\Psi$ itself is negligible due to its large mass. 

Let us compare the present parameter set to data obtained at SPS, which is in the same charged particle region as the pp data at LHC. 
To model the initial state for the $J/\Psi$-particles one needs the momentum distribution and the spatial distribution at the production points.\\
For the momentum distribution of the produced charmonia at SPS we use an ansatz from \cite{Linnyk:2008hp}. 
\begin{equation}
 \frac{dN}{dx_Fdp_T}\sim (1-|x_F|)^c\,e^{-b_{p_T}p_T}
\end{equation}
The distribution in $x_F$ is taken from \cite{Abramov:1991tn} with $c=a/(1+b/\sqrt s)$, where $a=16$ and $b=24.9$ \cite{Cassing:2000vx,Bratkovskaya:2003ux}. 
The exponent for the transverse distribution is $b_{p_T}=2.08\,$GeV$^{-1}$ \cite{Linnyk:2008hp}.\\
Our spatial distribution for the charmonium production at SPS is based on the Glauber model and calculated using the UrQMD model. 
Here we perform a pre-run for each event where we write down the nucleon collision points that we get when we switch off all interactions. 
These nucleon collision points are used in the sub-following full event for possible charmonium production coordinates.
Fig. \ref{SPS} depicts $J/\Psi$ decays to muons scaled by the number of Drell-Yan pairs dependent on the number of participants $N_{part}$ and the corresponding number of charged particles $N_{ch}$. 
UrQMD data are compared to data of the NA50 experiment \cite{Alessandro:2004ap}. In line with the expectations we observe a strong decrease of the $J/\Psi$ yield  towards central interactions. One should note that the present set-up is slightly different from the one suggested by Spieles et al. \cite{Spieles:1999kp,Spieles:1998pa}. Here we assume the presence of prehadronic states in comparison to the previous analysis that incorporated a time dependent formation time. There are also further differences in the energy dependence of the hadronic cross sections, which were assumed to be constant in the Spieles analysis.  

Let us now turn to the analysis of the LHC pp data. The momentum distribution of $J/\Psi$s at LHC has been measured with the various experiments at a center of mass energy of $7\,$TeV. Because we are interested in the late stage effects, we do not attempt to model the initial state, but fit the rapidity distribution of \cite{Aamodt:2011gj} using a double Gaussian fit 
\begin{equation}
\frac{d\sigma_{J/\Psi}}{dy}=\frac{A}{\sqrt{2\pi}\cdot\sigma}\cdot\left(e^{-\frac{1}{2}\left(\frac{y-\mu}{\sigma}\right)^2}+e^{-\frac{1}{2}\left(\frac{y+\mu}{\sigma}\right)^2}\right)
\end{equation}
and the transverse momentum distribution 
of \cite{Khachatryan:2010yr} and \cite{Aaij:2011jh} using 
\begin{equation}
\frac{d^2\sigma(J/\Psi)}{dp_Tdy}=p_T\cdot\left(A\cdot\left(1+\left(\frac{p_T}{B}\right)^C\right)\right)^{-z}
\end{equation}
with the parameters A,B,C and z. We fit the parameters for the transverse momentum distribution for each measured rapidity bin separately. The parameters are displayed in Table \ref{coefficients}. 

\newcolumntype{C}[1]{>{\centering}m{#1}}
\newcolumntype{R}[1]{>{\centering\arraybackslash}p{#1}} 

\newcommand{\ctab}{\centering\arraybackslash} 
\begin{table}[h]
\begin{tabular}{| C{2.5cm} | C{1cm} | C{1cm} | C{1cm} | R{1cm} |}
\hline
  rapidity & A & B & C & z \\
\hline
  $|y| < 1.2$ & 0.459 & 6.72 & 1.8 & 5.04 \\
  $1.2 < |y| < 1.6$ & 0.601 & 9.13 & 1.8 & 8.03 \\
  $1.6 < |y| < 2.4$ & 0.33 & 4.57 & 1.8 & 4.52 \\
  $2.4 < |y| < 2.5$ & 0.685 & 22.0 & 1.1 & 21.1 \\
  $2.5 < |y| < 3.0$ & 0.234 & 5.13 & 1.7 & 5.18 \\
  $3.0 < |y| < 3.5$ & 0.188 & 4.42 & 1.8 & 4.44 \\
  $3.5 < |y| < 4.0$ & 0.215 & 4.54 & 1.8 & 4.68 \\
  $|y| > 4.0$ & 0.282 & 4.95 & 1.75 & 5.46 \\
\hline
 \end{tabular}
\caption{Coefficients for the momentum distribution in Formula 5.}
\label{coefficients}
\end{table}
  
\begin{figure}
{\includegraphics[width=0.5\textwidth]{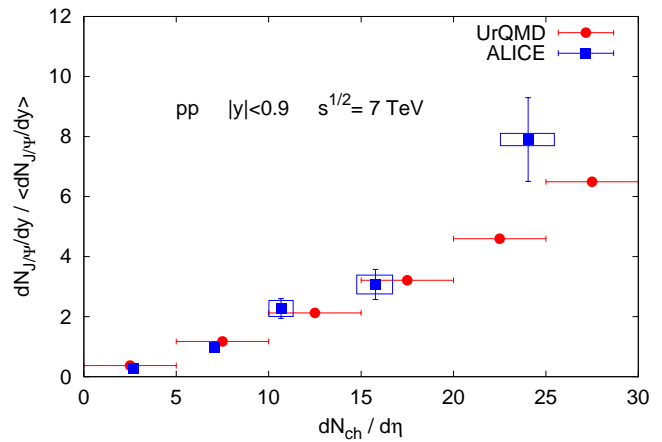}}
\caption{(Color online) Number of $J/\Psi$ particles dependent on the charged particle multiplicity in proton-proton collisions. 
UrQMD calculations are compared to measurements at ALICE \cite{Abelev:2012rz}. 
We employed a rapidity cut of $|y|<0.9$ for the $J/\Psi$s and $\eta <1$ for the charged particles to make our simulation comparable to the ALICE measurements.}
\label{LHCpp1}
\end{figure}

\begin{figure}
{\includegraphics[width=0.5\textwidth]{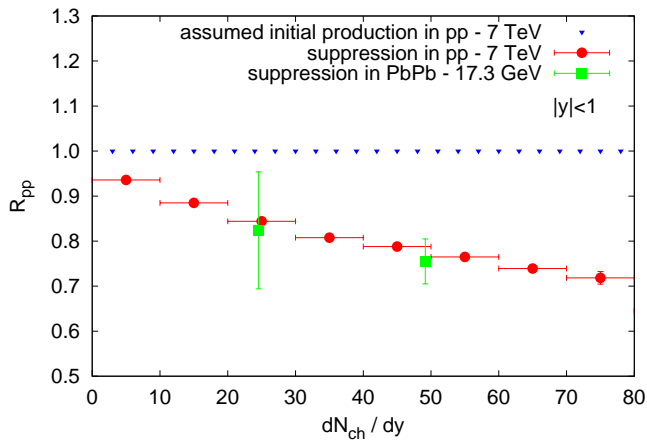}}
\caption{(Color online) Number of $J/\Psi$ particles dependent on the charged particle multiplicity. 
A rapidity cut of $|y|<1$ is employed for all particles. 
The $J/\Psi$ suppression in pp collisions at LHC and the Pb-Pb collisions at SPS show a very similar dependence on the charged particle multiplicity.}
\label{LHCpp2}
\end{figure}

To obtain the initial spatial distribution of the $J/\Psi$ emission points in pp reactions we smear the production points located in the proton-proton collision plane with a Gaussian distribution of variance $\sigma=0.88\,\textrm{fm}$.
Let us now explore pp collisions at a center of mass energy of $\sqrt {s_{NN}}=7\,$TeV within this approach. 
Since PYTHIA without tunes does not provide a satisfactory description of the charged particle vs. $J/\Psi$ yield correlation, 
we employ the $\frac{dN_{J/\Psi}}{dy}$ vs $N_{ch}$ distribution from \cite{Ferreiro:2012fb} for the initial $J/\Psi$ production. 
We now investigate the influence of the 'medium' created by analysis of the pp collisions in various charged particle multiplicity bins. 
Fig. \ref{LHCpp1} shows a comparison of our calculation to data measured at LHC. 
At low multiplicities one observes a good agreement to data. At higher charged particle multiplicities we observe a slight suppression of the $J/\Psi$ yield compared to the data. However, one should note that the magnitude of the input by \cite{Ferreiro:2012fb} has been tuned to describe the pp data without additional late stage effects, meaning that a slight re-adjustment of the parameters might improve the description of the data. 
To make the $J/\Psi$ suppression in high multiplicity pp reactions more visible we show the ratio $R_{pp}$ of finally observed $J/\Psi$s over initially produced $J/\Psi$s in Fig.\ref{LHCpp2} for each multiplicity bin at mid rapidity ($|y|\le 1$):
\begin{equation}
R_{pp}=\frac{dN_{J/\Psi}^{final}/dy |_{|y|\le 1}}{dN_{J/\Psi}^{initial}/dy |_{|y|\le 1}}
\end{equation}
At very high charged particle multiplicities in pp the $J/\Psi$s might be suppressed by $\sim 20\%$ as compared to the initially produced $J/\Psi$s at this multiplicity due to the interaction with the medium formed in the pp collision. Without a selection of charged particle multiplicity bins, i.e. minimum bias, the average suppression is on the order of 10\% . In the present set-up for the simulation, the prehadronic interactions, i.e. $J/\Psi$ melting and $J/\Psi$ dissociation by prehadrons, provide the major source of the suppression in pp collisions. The following hadronic phase has little effect on the $J/\Psi$ abundance. 
Let us also compare our results to the $J/\Psi$ suppression in Pb-Pb collisions at SPS energies at similar charged particle densities (Fig. \ref{LHCpp2}, squares). The similarity suggests that the $J/\Psi$ suppression may mainly depend on the charged particle multiplicity but not on the collision energy - at least if $J/\Psi$ recombination can be neglected. 

To summarize, we have explored the potential for charmonium suppression in inelastic proton-proton
collisions at $\sqrt {s_{NN}}=7$~TeV as function of the charged particle multiplicity. To this aim we have supplemented the UrQMD simulations with a phenomenological pre-hadronic stage to allow for $J/\Psi$ dissociation and melting in the hot and dense medium. 
It turns out that a small amount of $J/\Psi$ suppression (up to $20-30\%$ at $dN_{ch}/dy|_{|y|<1}>$ 70) may be encountered even in pp interactions at LHC.

\section{ACKNOWLEDGMENTS}
We are grateful to the Center for Scientific Computing (CSC) at Frankfurt for the computing resources.
T.Lang gratefully acknowledges support from the Helmholtz Research School on Quark Matter Studies.
This work was supported by the Hessian LOEWE initiative through the Helmholtz International Center for FAIR (HIC for FAIR).


\end{document}